\documentstyle[prl,aps,epsf]{revtex}
\draft
\begin{document}
\twocolumn[\hsize\textwidth\columnwidth\hsize\csname @twocolumnfalse\endcsname

\title{Directional Locking Effects and Dynamics for 
Particles Driven Through a Colloidal Lattice} 
\author{C. Reichhardt and C.J. Olson Reichhardt } 
\address{ 
Center for Nonlinear Studies and Theoretical Division, 
Los Alamos National Laboratory, Los Alamos, New Mexico 87545}

\date{\today}
\maketitle
\begin{abstract}
We examine the dynamics of a single colloidal particle driven through  
a colloidal lattice which can distort in response to the driven particle. 
We find a
remarkably rich variety of dynamical 
locking phenomena as we vary the angle of the applied drive
with respect to the orientation of the colloidal lattice. 
When the driven colloid locks to  
certain lattice symmetry directions, its motion is
not necessarily aligned with the drive.    
Applying a transverse force to the driven particle can result in
either increased or decreased drag in the driving direction,
depending on the angle of the drive.
The dynamical locking 
produces anomalies in both the longitudinal and the transverse velocity vs 
driving force curves, 
including steps and regimes of negative differential resistance.  
As the interaction of the driven particle with
the surrounding lattice increases, 
significant distortion or dislocations in the surrounding 
media occur, and as a result the directional locking is
enhanced.
We compare these results to those obtained for driving particles
over {\it fixed} substrates, and show
that a far richer variety of behaviors occurs
when the underlying lattice is allowed to distort.
We discuss how this system can be used for particle
species segregation when the onset of different locking angles 
occurs at different drives for varied particle sizes. 
We also show that the most pronounced locking phases should be
observable 
at temperatures up to the melting transition of the colloidal lattice. 
\end{abstract}
\pacs{PACS numbers: 82.70.Dd,05.45.-a}

\vskip2pc]
\narrowtext
\section{Introduction} 

Assemblies of 
interacting colloidal particles exhibit a wide range
of ordered, glassy, and liquid states \cite{Grier1,Lowen2}, and
have proven an ideal system for studying these equilibrium states with various
imaging techniques. 
Colloids can also be studied in nonequilibrium conditions 
such as under shear or when
driven with different types of fields \cite{Lowen2}.   
Recently Korda {\it et al.} \cite{Korda3} investigated the dynamics of moving
colloids driven over a square two-dimensional (2D) 
optical tweezer array. Here, the 
optical tweezers act as attractive sites \cite{lump3a}. In this
system, a series of locked transport states occur as the  
angle at which the colloids 
are driven with respect to the 
optical lattice is varied. In the locked states, 
the colloids move along a high-symmetry direction of the lattice 
even when this is not the same as the direction of drive.
The locking produces steps 
in the colloidal velocity vs driving angle,
and may have applications  
as a novel method for continuous separation
of particles of different species, 
such as macromolecules and biological
cells \cite{Kord4,Gopin5,GKorda6}. 
Separation can be achieved when a particular species
locks to a symmetry direction while the remaining species continue
to move in the direction of the applied drive.
Other techniques 
to separate particle species along this line
include the transport of 
particles through periodic obstacle arrays 
\cite{Chou7,Viovy8}.  

Particle transport through fixed periodic arrays under
varying drive direction has also been
investigated in  
simulations of vortices moving in 
nanostructured superconductors \cite{Reichhardt9} 
and Josephson junctions \cite{Stroud10,Candia11,Marconi12}. 
Recent experiments on vortices driven at
different angles through periodic pinning arrays 
find guiding effects similar to the features seen in 
simulations \cite{Look13}. 
Similar effects have also been studied  
in electron transport through periodic 
anti-dot arrays \cite{Ahn14} as well as atomic friction over
ordered surfaces \cite{Granato15}. In general,
for all of these systems,
the particle motion locks to certain symmetry angles of the
lattice.
Refs.~\cite{Reichhardt9,Ahn14} showed that 
the velocity vs angle curve has a Devil's staircase structure,
with a step in the particle velocity at each locking angle.
Certain highly symmetrical angles of drive
produce stronger locking than others.
       
In the experiments of Ref.~\cite{Korda3} and 
the theoretical studies of 
Refs.~\cite{Reichhardt9,Stroud10,Candia11,Marconi12,Ahn14,Granato15},  
the periodic lattice 
through which the particles move
is completely {\it rigid}. Another approach which has not been
previously considered is for the driven    
particle to move through a stationary, but {\it distortable},
lattice of particles.
In this case,
the driven particle still interacts with a periodic array
of obstacles; however, the driven particle can in turn affect the
surrounding media 
and distort or create dislocations in the surrounding lattice. 
A physical example
of such a system
would be driving a single or small number of colloidal
particles through a triangular
colloidal lattice. This can be achieved by placing a colloidal particle
that responds to an external drive 
into a colloidal lattice for densities and temperatures 
where the non-driven colloids form a triangular 
lattice. It should be straightforward 
to drive individual colloids with an optical tweezer \cite{GKorda6}. 
It is 
also possible to use magnetic impurity particles 
driven with a  magnetic field, as in recent experiments
by Weeks {\it et al.} \cite{Weeks16}.
A variation
on this type of system would be to have one colloid or impurity 
particle fixed at a spot and to drive the colloidal lattice past it. 
Experiments 
along this line have been performed 
previously for large obstacles \cite{Weiss17}.     

A system that has many similarities to colloids is vortices in
superconductors, where the mutual repulsion between vortices leads to
the formation of a triangular lattice. 
For a clean superconductor where the 
intrinsic pinning is not strong enough to distort
the vortex lattice, 
it may be possible to drive a single vortex
with an MFM or STM tip. 
It has also been demonstrated experimentally using
a lithographically created artificial pin
\cite{Matsuda18}  that 
a single pinning site can capture one or more 
vortices while the other vortices are driven past.  
Other similar systems include 
driving an obstacle through a triangular bubble raft lattice,
or manipulating a single droplet of a two ferrofluid system in
a triangular droplet state.
Another system closely related to colloids is
2D dusty
plasmas where charged dust particles form a triangular lattice \cite{Goree19}
and individual particles can be driven with lasers. 

In a system where the periodic lattice through which a particle is driven
can itself be affected by the driven particle, 
a number of intriguing 
scenarios are possible which cannot arise 
when the substrate is completely rigid.
Examples include a transition 
from an ordered flow, where the driven particle
only distorts the lattice elastically and does not create dislocations,
to a disordered flow, where the particle tears the lattice
and dislocations are generated. 
This could occur when the particle moves at an angle that is not
commensurate with the lattice, or for a large enough driven
particle.
Other effects that could occur include the possibility that,
for certain driving directions, 
the driven particle may drag portions of the
non-driven lattice with it.  Conversely, for other driving directions, 
the particle may be able
to slip freely between the particles comprising the lattice. 
Since the lattice can 
locally distort to more easily
accommodate the motion of the driven particle in the locked regimes, 
the locking effects may be different in nature
and more pronounced 
than those observed in previous work on rigid substrates.
One practical application of this system, 
as suggested in Refs.~\cite{Korda3,Kord4}, 
is the continuous separation of different species of
particles as they move through periodic arrays of obstacles.
We demonstrate that the separation 
of particles of different sizes 
is enhanced by the ability of the surrounding
lattice to distort.
If, under certain conditions,
the distortable lattice can
act as a filter for different sized particles,  
then the additional step of using optical tweezer arrays, 
as in Ref.~\cite{Korda3}, would not be
needed.

\section{Simulation}
We simulate a 2D monodisperse colloidal assembly 
using Langevin dynamics, with techniques that have
been described previously 
\cite{Olson20,Hastings21}. 
The overdamped equation of motion for 
a single colloid $i$ in a system with periodic boundaries in the
$x$ and $y$ directions is  
\begin{equation} 
{\eta} \frac{d {\bf r}_{i}}{dt} = {\bf f}_{ij} + {\bf f}_{d}^i+ {\bf f}_{T} \ .
\end{equation} 
Here
$\eta=1$ is the damping term arising from the
fact that the particle is moving through a viscous media.
The interaction force from the other colloids is
${\bf f}_{ij} = -\sum_{j \neq i}^{N}\nabla_i V(r_{ij})$, where 
the repulsive colloid-colloid interaction 
is Yukawa or screened Coulomb,
\begin{equation} 
V(r_{ij}) = \frac{q_{i}q_j}{|{\bf r}_{i} - {\bf r}_{j}|}
\exp(-\kappa|{\bf r}_{i} - {\bf r}_{j}|). 
\end{equation}
Here $q_{i(j)}$ is the charge of the particle, 
$1/\kappa$ is the screening length, ${\bf r}_{i}$ is the position of
particle $i$, and ${\bf r}_{j}$ is the position of particle $j$.   
We initialize the system in a triangular lattice with $N$ colloids
and a lattice constant $a$, and then insert one additional  
driven colloid.
The screening length is fixed at $1/\kappa=3a$. 

The thermal force ${\bf f}_{T}$ is a randomly fluctuating force 
modeled as random kicks, with
the properties 
$<{\bf f}_T(t)> = 0$ and 
$<{\bf f}_T(t){\bf f}_T(t^{\prime})> = 2\eta k_{B}T\delta(t -t^{\prime})$. 
For most of the results presented here, the colloid density
is held fixed and the system size is set to $L=34a$.
With these parameters there is a well defined melting temperature
$T_{m}$ for the unperturbed lattice, which 
we determine by measuring the fraction of colloids which are 
six-fold coordinated.
We measure temperature in units of $T_m$.
For the first part of the paper we consider the case of $T = 0$ 
so that the effects of the dynamics can be clearly distinguished. We 
later show that, for a large range of $T/T_{m}$, 
essentially the same results appear, 
and that the locking effects are lost for $T/T_{m} > 1$. 
Thus the results from our work should be applicable to experimental
samples at finite temperatures.    

We vary the charge on the driven particle $q_{d}$ and the direction of drive.
All particles except the driven one have charge $q=1$ and applied dc drive
${\bf f}^{i}_d=0$.
The force on the driven particle has two 
components:
\begin{equation}  
{\bf f}_{d} = f^{d}_{x}{\bf {\hat x}} + f^{d}_{y}{\bf {\hat y}}. 
\end{equation} 
We hold $f_{x}^d$ fixed at a constant value and increase $f_{y}^d$. 
We then monitor the velocities $V_{x}$ and $V_{y}$ of the
driven particle.  

We do not take
into account possible 
hydrodynamic effects or possible long-range attractions between
colloids. 
We have conducted simulations for different system sizes, lattice
constants $a$, 
and screening lengths, and find that the qualitative results 
presented here 
are robust. 

\section{Small Driven Particle}
We first consider the case where the driven particle interacts only
weakly with the surrounding particles so 

\begin{figure}
\center{
\epsfxsize=3.5in
\epsfbox{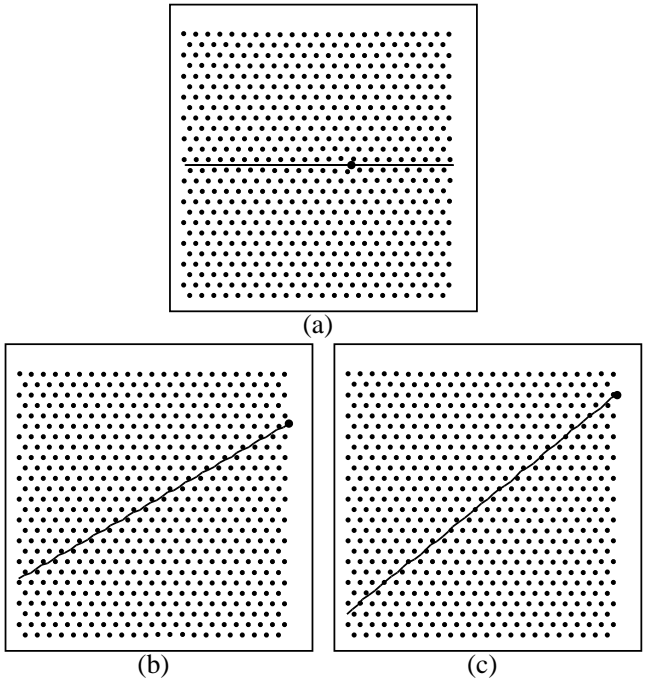}}
\caption{
Particles (black dots) and trajectories (black lines)
for a system with $q_{d}/q = 0.35$ at
$f_{x}^d=1.0$ and:
(a) $f^{d}_{y}/f^{d}_{x} = 0.0$,   
(b) $0.57$, and (c) $0.78$.  
}
\end{figure}

\noindent
that, although small distortions of the lattice can 
occur, no dislocations are formed in the surrounding media. 
We define this weak interaction regime as $q_{d}/q < 0.5$. 
In the case of a triangular lattice 
aligned in the $x$-direction,
as in Fig.~1(a), 
locking should occur for angles of drive where the driven colloid
can move freely between the colloids in the lattice. This 
occurs for ratios of the applied drive of 
\begin{equation}
\frac{f_{y}^d}{f_{x}^d} = \frac{{\sqrt 3}m}{2n + 1} 
\end{equation} 
where $m$ and $n$ are integers. The corresponding 
locking angle is
$\theta = \arctan(f_{y}^d/f_{x}^d)$.
Eq. 4 predicts locking at $60^{\circ}$ 
corresponding to $m = 1$, $n = 0$; 
$30^{\circ}$ corresponding to $m = 1$, $n = 1$; $19.1^{\circ}$
corresponding to 
$m = 1$, $n = 2$; and $0^{\circ}$ for $m = 0$. 
Other locking angles occur for higher order values of $m$ and $n$.  
In Fig.~1 we illustrate the driven particle,
the surrounding lattice, and the particle trajectories
for a system with fixed parameters of
$q_{d}/q = 0.35$ 
and $f_{x}^{d} = 1.0$ 
for increasing 
$f_y^d/f_x^d=0,$ 0.57, and 0.78. The trajectories
in all these cases show that 
the driven particle moves through the lattice without significantly
distorting it, due to the weak charge of the driven particle.
In Fig.~1(a) we show the commensurate 
case of $f_{y}^{d}/f_{x}^{y} = 0$
where the particle channels between the surrounding colloids. 
Fig.~1(b) presents another commensurate case for 
$f_{y}^{d}/f_{x}^{d} = 0.57$ where the 
particle moves in a periodic orbit 
along a symmetry line of the lattice.
This orbit corresponds to $m = 1$, $n = 1$ from Eq.~4, so the
flow is along $30^{\circ}$. 
In Fig.~1(c) we show an 
incommensurate case of $f_{y}^{d}/f_{x}^{y} = 0.78$, which is not 
a simple

\begin{figure}
\center{
\epsfxsize=3.0in
\epsfbox{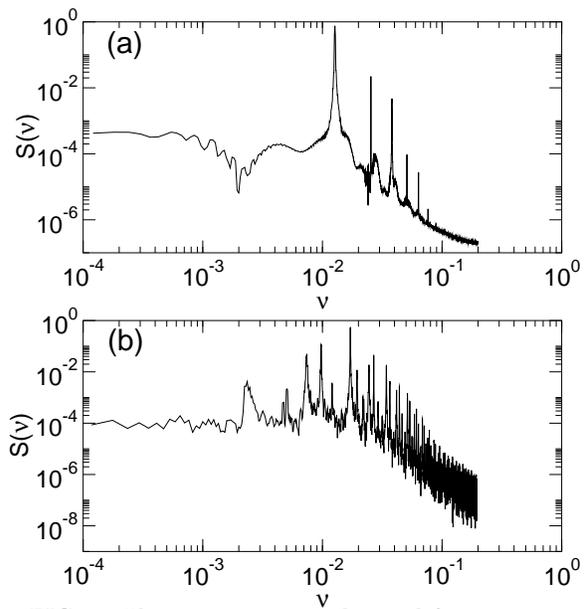}}
\caption{
The power spectra obtained from a time series
of $V_{y}$ for: (a) 
the commensurate case $f_y^d/f_x^d=0.57$ shown in Fig.~1(b), and (b)
the incommensurate case $f_y^d/f_x^d=0.78$ shown in Fig.~1(c). 
}
\end{figure}

\noindent
ratio in Eq.~4. 
Here the particle does not move in a completely periodic 
orbit 
but shows some non-periodic deviations; however, the 
particles in the surrounding lattice
still show little deviation from their positions. 
We observe similar types of motion at the other incommensurate
angles as well.

In order to better demonstrate the difference 
between the commensurate and
incommensurate particle orbits in Fig.~1, 
we analyze the power spectra of the velocity $V_y$ of the 
driven particle for the two cases.  
In Fig.~2 we show the
power spectra $S(\nu)$ 
obtained from a time series
of $V_{y}$ of the driven particle at fixed
$f^{d}_{y}/f^{d}_{x}$.
Figure 2(a) illustrates $S(\nu)$
for the commensurate case $f_y^d/f_x^d=0.57$ shown in Fig.~1(b). 
Here the spectrum has
a prominent peak at $\nu=0.0127$ inverse molecular dynamics steps 
with higher harmonics, 
indicating that the particle motion
is strictly periodic with a single frequency.
In a commensurate orbit, a single frequency is expected
to appear since
the particle is moving along a symmetry direction of the lattice
and is slowed by the interactions from the surrounding
lattice at a constant spacing and a constant rate. 
In Fig.~2(b), the spectrum 
for the incommensurate case $f_y^d/f_x^d=0.78$ shown
in Fig.~1(c) has many peaks,
indicating that the particle
is undergoing motion with many different periods giving 
a broad spectrum.  
In general, for these incommensurate phases, the particle likely jumps
between several different 
closely spaced locked phases with high $m$ and $n$ values which,
together, produce an angle of motion which is close to the angle of the drive.
We find similar behaviors
in the power spectra
at other commensurate and incommensurate
angles.

In Fig.~3(a) we
simultaneously plot $V_{x}$ and $V_{y}$ 
vs $f^{d}_{y}/f^{d}_{x}$, 

\begin{figure}
\center{
\epsfxsize=3.5in
\epsfbox{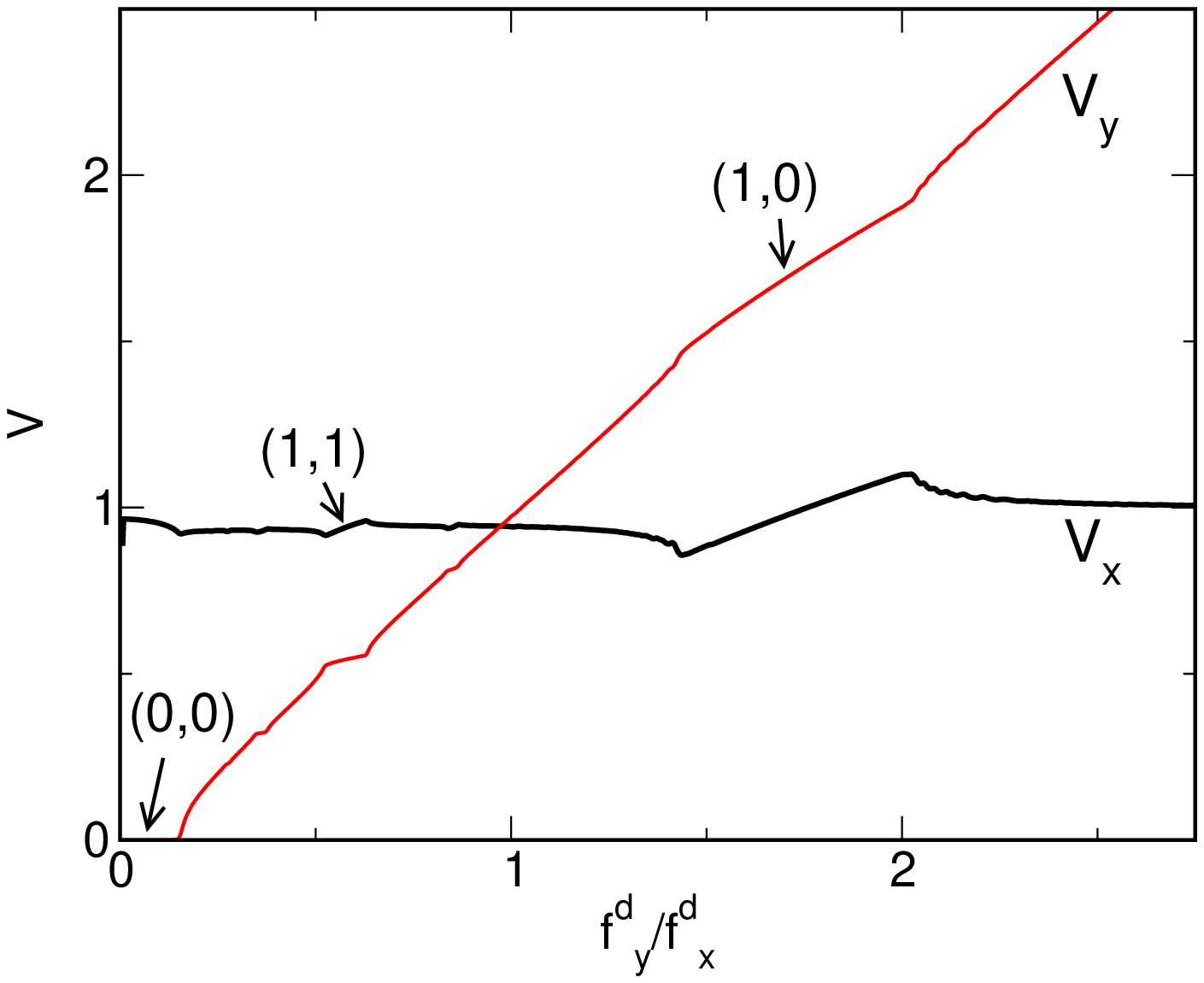}
\epsfxsize=3.5in
\epsfbox{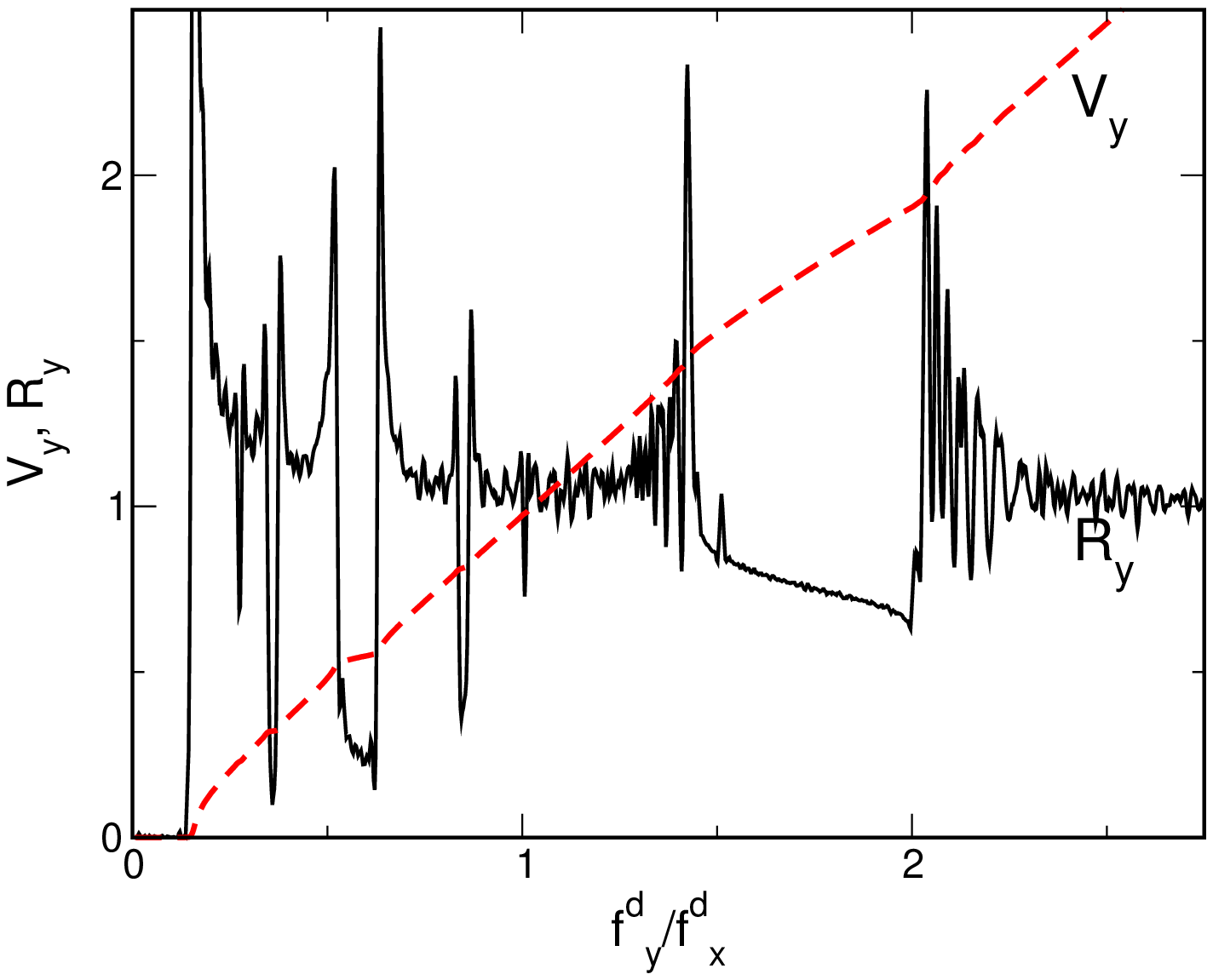}}
\caption{
(a) $V_{x}$ (dark line) and $V_{y}$ (light line) for 
fixed $f^{d}_{x} = 1.0$ and increasing $f^{d}_{y}$
for $q_{d}/q = 0.35$. 
(b) $V_y$ (dashed line) and its derivative $R_y$ (solid line), showing the
locking phases more clearly.
}
\end{figure}

\noindent
with fixed $f_{x}^d = 1.0$, showing that the locked phases strongly affect the
transport proprieties.
We highlight a few ratios of $(m,n)$ from 
Eq.~4 to indicate the most prominent locking phases:
$(0,0)$, which starts at $f_{y}^d/f_{x}^d = 0$
in the bottom left corner of Fig.~3(a);
$(1,1)$ for flow at $30^{\circ}$; and the very prominent
$(1,0)$ locking centered at $f_y^d/f_x^d=1.7$. There are additional smaller
steps that are difficult to see for this value of $q_{d}$.  
Another interesting feature is that $V_{x}$
has an average value of $0.9$
for $f_{y}^d/f_{x}^d < 2.0$.
We note that, for a single 
driven colloid moving in the absence of any other particles,
we would have $V_{x} = 1.0$, which is higher 
than the value obtained in the presence of the other particles.
This indicates 
that the driven particle experiences an additional damping 
due to the interactions with the
other particles. This damping originates when the surrounding colloids 
shift slightly as the driven particle moves past.
Since the motion of these surrounding colloids
is also overdamped, the only source of energy
is the driven particle.   
In the case of a rigid substrate \cite{Reichhardt9}, this
additional damping effect is absent 
since the periodic substrate cannot distort
and absorb energy.
Another immediately obvious difference between this system and
particle motion over a rigid substrate
is that in Fig.~3(a), none of the
steps in $V_{y}$ have $dV_{y}/df^{d}_{y} = 0$ 
\cite{Grier1,Reichhardt9,Ahn14}.
Although the slope of $V_y$ is reduced on the steps, it does not
drop to zero,
indicating that along the
locking directions the particle velocity $V_y$ continues to increase 
with $f_{y}^d$.     
Fig.~3(a) also indicates that,
in the locking regions, $V_{x}$ is not constant but shows 
features where the
velocity both {\it increases} and {\it decreases} 
which coincide with slope changes 
in $V_{y}$. 
On the steps, when $V_{y}$ locks to the lower slope value,
$V_{x}$ locks 
to a constant positive slope which can be seen most clearly near 
$f_{y}^d/f_{x}^d = 1.7$ at the $60^{\circ}$ locking of $(1,0)$. 
At the lower edge of the $(1,1)$ and $(1,0)$ steps,
$V_x$ {\it decreases} in an abrupt dip before rising linearly on the step.
On the upper end of these steps, when the slope in $V_{y}$ increases again, 
$V_{x}$ shows another decrease.

In Fig.~3(b) we plot $V_y$ along with its derivative $R_y$ to show
the locking phases more clearly.  At the $(1,0)$ locking, $R_y$
drops to a lower value due to the smaller slope of $V_y$ in the
locking regions.  Peaks appear in $R_y$ at the beginning and end of
each locked phase, reflecting the increased slope of $V_y$ outside
of the locked regions.  $R_y$ also shows some oscillations above
the $(1,0)$ locking, which we discuss in more detail in 
Section IV-C.  Additionally, locking phases appear as dips in
$R_y$ for the $(1,1)$ locking centered at
$f_y^d/f_x^d=0.57$ as well as at $(1,2)$ and $(1,3)$.

Fig.~3(a)  also demonstrates
that there is a clear {\it transverse depinning} threshold 
for the $(0,0)$ locking: 
$V_{y} = 0.0$ for $ f_{y}^d/f_{x}^d < 0.15$.
Here, despite the fact that the particle is moving
in the $x$-direction, there is an effective
pinning threshold for motion in the $y$-direction. 
Below this depinning threshold,
the particle trajectories 
are the same as those shown in Fig.~1(a), where the particle
moves in an effective 1D trough. 
Transverse depinning thresholds
have been observed in various
systems of particles driven longitudinally over a 2D periodic
substrate as an increasing transverse force is applied.
These systems include
longitudinally moving vortices in periodic pinning arrays 
\cite{Reichhardt9,Stroud10,Marconi12} 
and atoms moving over periodic surfaces \cite{Granato15}. 
In colloidal experiments on periodic substrates,
a locking is observed for small angles \cite{Korda3}. 
In all these cases the 
particle motion remains locked in the longitudinal direction until 
a large enough transverse force is applied.  
In Fig.~3(a) as $f_{y}^d/f_{x}^d$ approaches 
the transverse depinning threshold,  
$V_x$, the particle velocity in the $x$-direction, decreases.
This decrease occurs because the increasing transverse force shifts the
driven particle closer to the lattice particles, 
increasing the interaction between the 
driven particle and the other particles.
Since the surrounding lattice is flexible, some of the 
motion of the driven particle  
is transfered into small 

\begin{figure}
\center{
\epsfxsize=3.5in
\epsfbox{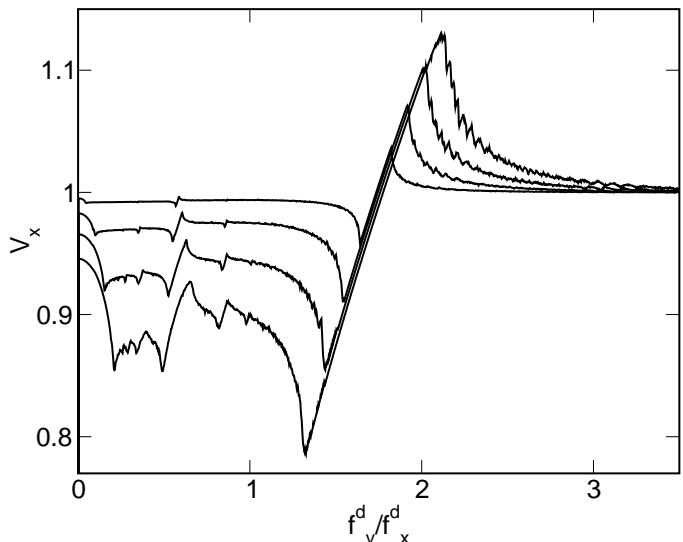}}
\caption{  
$V_{x}$ vs $f^{d}_{y}/f^{d}_{x}$ for $q_{d}/q = 0.125$, 0.25, 0.35,
and $0.5$ (from top left to bottom left). 
}
\end{figure}

\noindent
distortions of the 
surrounding lattice.  These distortions increase 
as the driven particle is forced closer to the particles in the lattice.
Once the driven particle begins to move in the $y$-direction, its
velocity in the $x$-direction increases
since the driven particle 
no longer approaches the particles in the surrounding lattice
as closely and thus experiences less damping of its motion.
Remarkably, for 
$1.8 < f_{y}^d/f_{x}^d < 2.0$,
$V_{x}$ is {\it greater} than $1.0$, indicating that the 
particle is moving {\it faster} than 
it should given the applied driving force in the
$x$ direction. 
This indicates that some of the energy from the $y$ 
component of the drive is being  
transferred to the $x$-direction. 
Above the $(1,0)$ step for $f_{y}^d/f_{x}^d > 2.0$,
$V_{x}$ decreases slowly  back to $1.0$. 

We next consider how the locking regions evolve with
increasing $q_{d}$.
In Fig.~4 we plot $V_{x}$ vs $f^{d}_{y}/f^{d}_{x}$ 
for $q_{d}/q = 0.125$, 0.25, 0.35, and $0.5$
with $f_x^d=1.0$.
The plots are overlaid with no offset.
For $f_{y}^d/f_{x}^d = 0$, $V_{x}$ decreases 
for increasing $q_{d}$, indicating that the damping increases
as $q_d$ rises.
This decrease in $V_x$ occurs because 
the driven particle interacts
more strongly with the surrounding particles, causing larger distortions
and transferring more of its energy into
the lattice.
Additionally, all of the locking phases 
become wider with increasing $q_{d}$, 
indicating that as the distortions in the surrounding lattice
from the driven particle increase,
the channeling effect becomes stronger.
The locking phases for the higher order values of 
$(m,n)$ can now be clearly
resolved, especially for the $q_{d}/q = 0.5$ curve where a 
particularly large number of dips 
occur for $f^{d}_{y}/f^{d}_{x} < 1.0$. 
For the locking at $(1,0)$, the value of $V_{x}$ exceeds 1.0 
for $f^{d}_{y}/f^{d}_{x} > 1.7$ at all values of $q_d$. 
Just above the $(1,0)$ locking, $V_{x}$ decays
back to 1.0.  During this decay, there are clear small
scale periodic peaks which are a real effect 
that we study later in this paper. 

We note that for small $q_d$, the behavior of this system
is similar to that of a particle moving through a rigid 

\begin{figure}
\center{
\epsfxsize=3.5in
\epsfbox{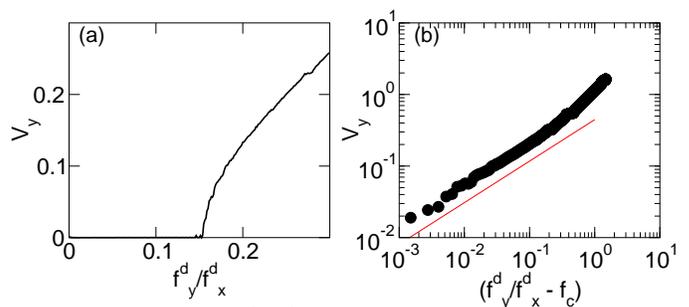}}
\caption{  
(a) $V_{y}$ vs $f^{d}_{y}/f^{d}_{x}$ 
at the transverse depinning 
transition for the system shown in Fig.~3 with
$q_{d}/q = 0.35$. (b) Log-log plot of $V_{y}$ vs
$(f^{d}_{y}/f^{d}_{x} - f_{c})$ where $f_{c}=0.15$ is the
transverse depinning threshold. The solid line is a power
law fit with an exponent of $\beta = 0.58$. 
}
\end{figure}

\noindent
lattice in Ref.~\cite{Reichhardt9}, where the
same type of anomalous steps were observed in both  $V_{x}$ and $V_{y}$. 
The locations of the steps in 
Ref.~\cite{Reichhardt9} occurred for different
values of $f^{d}_{y}/f^{d}_{x}$ since the underlying lattice was square rather
than triangular. 
Additionally, the slope of $V_{y}$ along the steps was nearly
zero in the rigid lattice,
making $V_y$ appear much more step-like. 
The similar behavior found here at small $q_d$
occurs when the driven particle
can no longer cause significant distortions in the surrounding
lattice, rendering it effectively rigid.

\subsection{Dynamics of Transverse Depinning}

We now examine the 
initial transverse depinning transition in more detail. We first consider
the shape of $V_{y}$ vs $f^{d}_{y}/f^{d}_{x}$ near 
the transverse depinning. For 
elastic media depinning in the presence
of quenched disorder, theory predicts  
$V \propto (f^{d} - f_{c})^{\beta}$, where $f_{c}$ is the
depinning threshold
\cite{Fisher22}. For a   
single particle depinning from a sinusoidal
substrate in 1D, $\beta = 1/2$. 
In 2D systems of collectively interacting particles where
plastic deformation occurs at depinning,
studies find scaling in the velocity
force curves with $\beta > 1.5$ \cite{Dominguez23}. 
It is not known if there would be any scaling for the transverse depinning
of a longitudinally moving particle.  
In Fig.~5(a) we show $V_{y}$
near the onset of motion in the $y$-direction for the 
system in Fig.~3 
with $q_{d}/q = 0.35$. We plot
$V_{y}$ vs $f^{d}_{y}/f^{d}_{x} - f_{c}$ on a log-log scale 
in Fig.~5(b), where $f_{c}=0.15$ is the 
transverse depinning force.
We find a reasonable power-law fit with $\beta = 0.58$
(solid line in Fig.~5(b)), close to the single 
particle value of $\beta=1/2$. 
This indicates that the transverse depinning is 
elastic in nature, and that when the particle depins it does not induce
tearing in the surrounding lattice. 

For the steps above $(0,0)$, whenever the particle leaves a locked region, 
$V_{y}$ shows curvature similar to that of the 
initial transverse depinning. 
This suggests that on the higher order steps, when the particle
is channeling along an easy flow direction, the particle is
effectively  pinned in the direction transverse to 
${\bf f}_{d}$, rather than in the direction 
transverse to $f^{d}_{x}$. 
As $f^{d}_{y}$ increases, the 
transverse force 

\begin{figure}
\center{
\epsfxsize=3.5in
\epsfbox{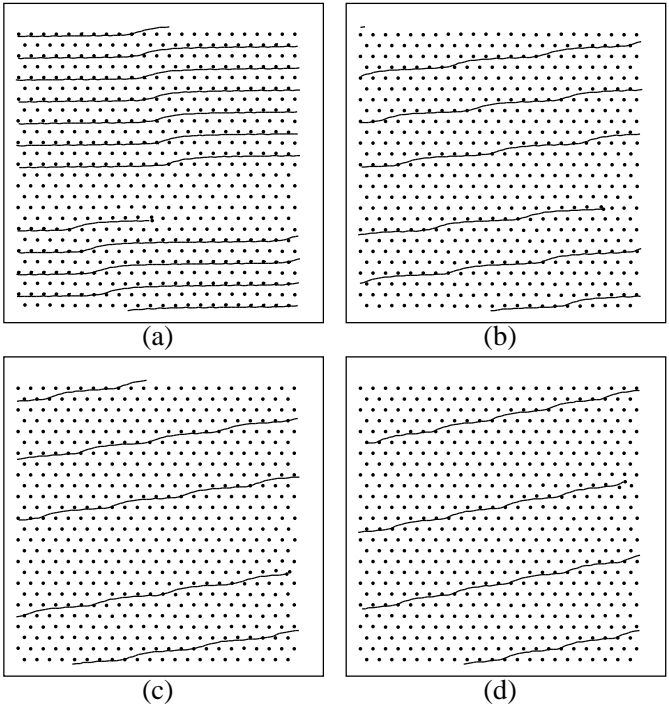}}
\caption{ 
Particles (black dots) and trajectories (black lines) near 
the transverse depinning transition for the system shown in Fig.~5 at 
$f^{d}_{y}/f_{c} =$ (a) 1.05, (b) 1.22, (c) 1.34,
and (d) 1.5. 
}
\end{figure}

\noindent
eventually becomes high enough
for the particle to depin and begin moving at a new angle. 

In Fig.~6 we plot
the particle trajectories at the transverse depinning
transition for the system shown in Fig.~5. 
In Fig.~6(a), just above the initial transverse depinning 
at $f^{d}_{y}/f_{c} = 1.05$, 
the particle stays locked along the $x$-direction 
for about $16a$ before moving 
in the positive $y$ direction
to the next row.
The particle moves in a 
staircase fashion. 
In Fig.~6(b) for $f^{d}_{y}/f_{c} = 1.22$, we find a staircase 
motion similar to that of Fig.~6(a),
but with the particle translating in the
$y$ direction every $7a$, 
producing a net $V_{y}$ that is higher than that of
Fig.~6(a). 
In Fig.~6(c) at $f^{d}_{y}/f_{c} = 1.34$
and Fig.~6(d) at $f^{d}_{y}/f_{c} = 1.5$, the staircase motion persists
with the particle moving shorter distances in the $x$-direction 
between jumps to the next row in the $y$ direction.

\section{Intermediate Sized Driven Particles}
Next we consider the case for $0.5 < q_{d}/q < 3.0$, which we term the
intermediate particle regime.
Here we observe that the locking phases become more pronounced
and appear more step-like in $V_{y}$ with $dV_{y}/df^{d}_{y} = 0$
in some places.
We also find regimes where $V_{y}$ decreases for increasing
$f_{y}^d$, indicating a {\it negative} 
differential transverse resistance, or $dV_{y}/df^{d}_{y} < 0$. The
scaling in the initial transverse depinning is also 
lost and is replaced by 

\begin{figure}
\center{
\epsfxsize=3.5in
\epsfbox{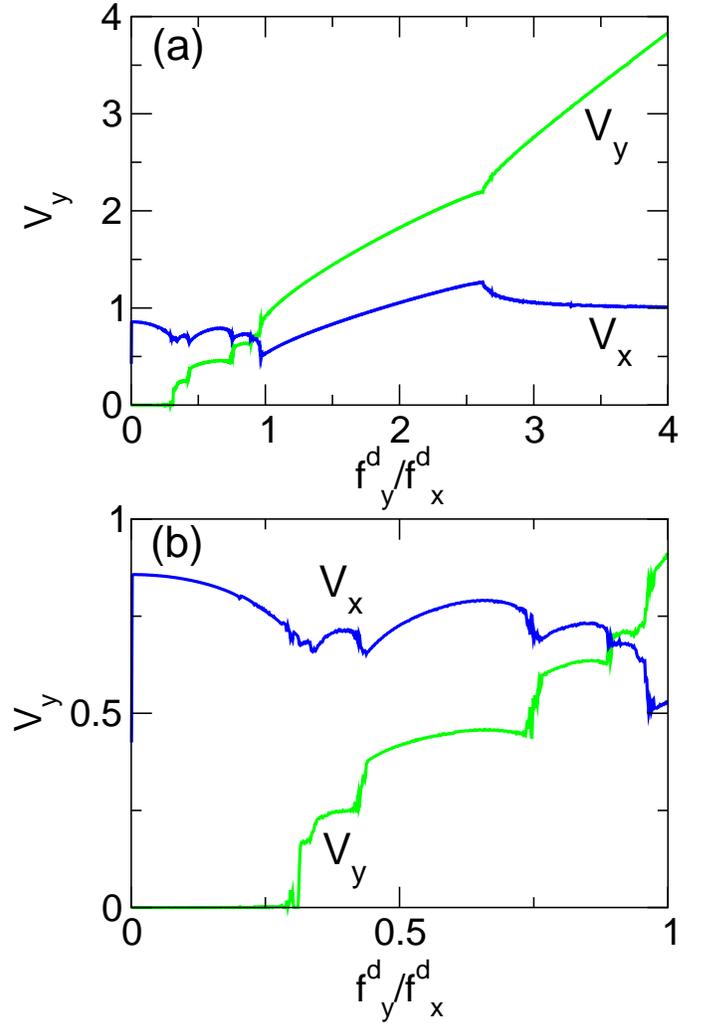}}
\caption{  
(a) $V_{x}$ (dark line) and $V_{y}$ (lighter line) vs $f^{d}_{y}/f^{d}_{x}$ 
for $q_{d}/q = 1.0$. (b) Closeup of (a) for 
$ 0.0 < f^{d}_{y}/f^{d}_{x} < 1.0$ 
}
\end{figure}

\noindent
a sharp jump, indicating that for $q_{d}/q > 0.5$ the system is no
longer in the elastic regime.  
In Fig.~7(a) we show $V_{x}$ and $V_{y}$ for 
a system with $q_{d}/q = 1.0$. 
For these intermediate $q_{d}$ values, the width of the
$(1,0)$ locking becomes so large that it begins to 
overlap with the
smaller locking phases, and some of the smaller locking steps are lost. 
The dips in $V_{x}$ are more pronounced  
and there are several regions 
where both
$V_{x}$ and $V_{y}$ show jumps into and out of different locked phases,
rather than the smoother transitions seen for lower values of $q_{d}$.  
The slopes of $V_{x}$ and $V_{y}$ along the $(1,0)$ locking 
are both positive but show a nonlinear bowing effect. 
In Fig.~7(b) we plot $V_{x}$ and $V_{y}$ 
for $0 < f^{d}_{y}/f^{d}_{x} < 1$, showing the smaller locking
phases more clearly. 
Along the step $(1,1)$, 
centered at $f_{y}^d/f_{x}^d = 0.6$, the
slope of $V_{y}$ shows a bow feature, indicating that the slope 
becomes {\it negative} near $f_{y}^d/f_{x}^d = 0.73$. 
On the next step we find a similar bow
feature.
This behavior is in contrast to the studies with rigid substrates, 
where the slope of $V_{y}$ never drops below zero along the 

\begin{figure}
\center{
\epsfxsize=3.5in
\epsfbox{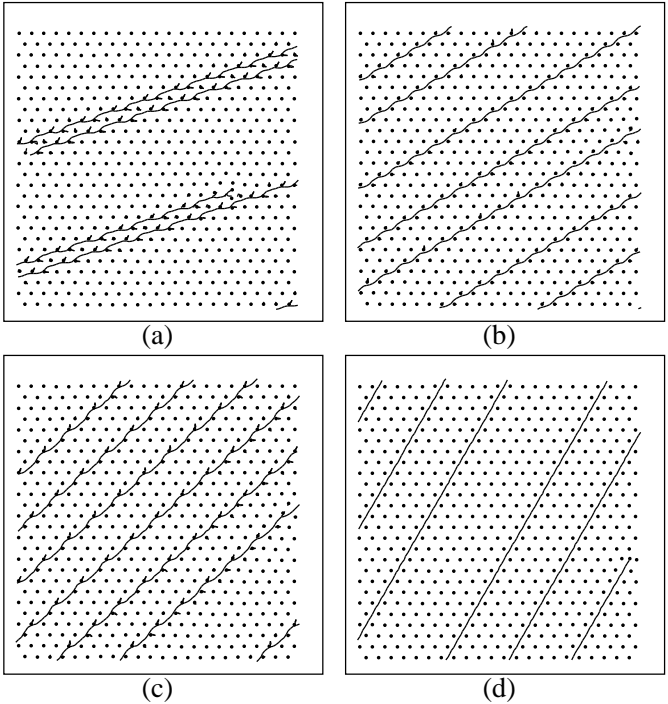}}
\caption{  
Particles (black dots) and trajectories (black lines) for the 
system in Fig.~7 with $f^{d}_{y}/f^{d}_{x} =$ (a) 0.39,
(b) 0.6, (c) 0.8, and (d) 2.0. 
}
\end{figure}

\noindent
steps. 
Fig.~7 also shows that both $V_{x}$ and $V_{y}$ 
can decrease simultaneously, indicating
an overall increase in the damping. For these 
intermediate $q_{d}$ values there 
is considerably more distortion in the surrounding lattice. Just before 
the driven particle exits a locked region, it has the strongest
interaction with the surrounding media and experiences the largest
damping.
At the jumps in velocity accompanying each step,
the driven particle can cause enough distortion 
to temporarily generate a localized dislocation in the lattice.

\subsection{Trajectories in the Locked Phases}

In Fig.~8 we 
illustrate some of the particle trajectories for different locking regimes
from the system in Fig.~7.
Fig.~8(a) shows the step near $f^{d}_{y}/f^{d}_{x} = 0.39$, 
which corresponds to the $m = 1$, $n = 2$ locking phase.  
The particle moves in 
a zig-zag pattern.  The 
elastic distortions in the surrounding colloid lattice are clearly visible
in the form of very small loops on both sides 
of the path of the driven particle. 
Here, the colloids closest to the driven particle
move a distance less than $a$  
as the driven particle approaches, and then return
to their initial positions after the driven
particle passes. There are some additional smaller distortions
in the lattice at larger distances from the driven colloid;
however, these motions are too small to be 
visible at the resolution in Fig.~8. 
In general, when the intermediate sized particles
move through a channel 

\begin{figure}
\center{
\epsfxsize=3.5in
\epsfbox{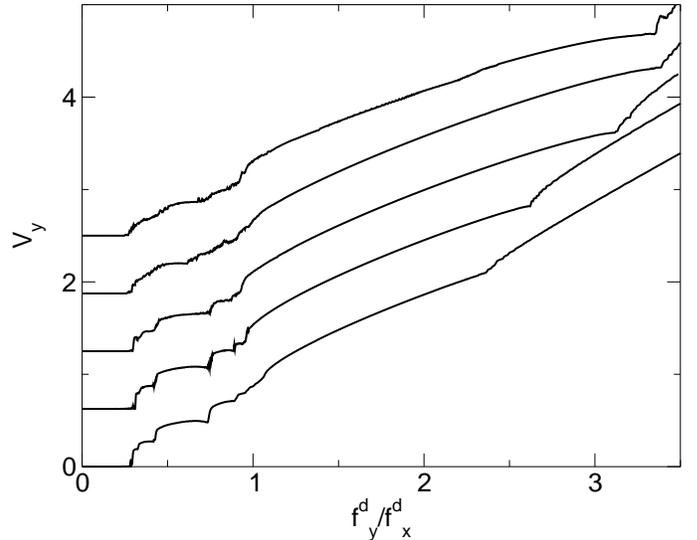}}
\caption{$V_{y}$ vs $f^{d}_{y}/f^{d}_{x}$ for varied $q_{d}/q$.
A systematic shift is added
in the $y$ direction for clarity. From top to bottom,
$q_{d}/q =$ 2.0, 1.75, 1.5, 1.0, and 0.75.   
}
\end{figure}

\noindent
during a locked phase, 
the motion is elastic and no dislocations are generated.
Only during the transitions out of the locked phases do
temporary distortions greater than $a$ in the surrounding lattice occur.

Fig.~8(b) shows the trajectories at $f_{y}^d/f_{x}^d = 0.6$, when
the particle moves at $30^{\circ}$ in the $(1,1)$ locking regime. 
The width of this $(1,1)$ 
locked region is considerably 
larger than that of the $(1,2)$ phase in Fig.~8(a).
Here, the particle follows a sinusoidal
path and causes distortions in the surrounding lattice; however,
these distortions are of a smaller magnitude than those in Fig.~8(a),
indicating that the particle channels more easily on the $(1,1)$ step. 
In Fig.~8(c) we plot the trajectories for the step
region at $f_{y}^d/f_{x}^d = 0.8$,
where more significant distortions in the
surrounding lattice occur than for Fig.~8(b). 
The width of this step is smaller than the $(1,1)$ step in 
Fig.~8(b).  
In Fig.~8(d) we show the trajectories at $f^{d}_{y}/f^{d}_{x} = 2.0$ 
for the most prominent step, $(1,0)$,   
where the vortices channel 
at $60^{\circ}$. 
There are almost no distortions in the surrounding lattice. 
For a value of $f^{d}_{y}/f^{d}_{x}$ just below the end of
the $(1,0)$ phase, $f^d_y/f^d_x \approx 2.6$, 
significant distortions of the colloids to the
upper left of the driven particle occur
in the surrounding lattice, 
since the driven particle is pushed in this
direction by the increasing $f^{d}_{y}$. 
When these distortions become large
enough, the $(1,0)$ phase ends.

\subsection{Locked Phase Evolution as a Function of $q_{d}$}

In Fig.~9 we plot $V_{y}$ vs $f_{y}^d/f_{x}^d$ for increasing values of $q_{d}$.
From bottom to top, $q_{d}/q =$  0.75, 1, 1.5, 1.75, and 2.0. In this 
range of $q_{d}$, the initial transverse depinning threshold changes little;
however, the width of the $(1,0)$ locking 
which begins at $f_{y}^d/f_{x}^d \approx 1.1$ 
grows until it saturates 

\begin{figure}
\center{
\epsfxsize=3.5in
\epsfbox{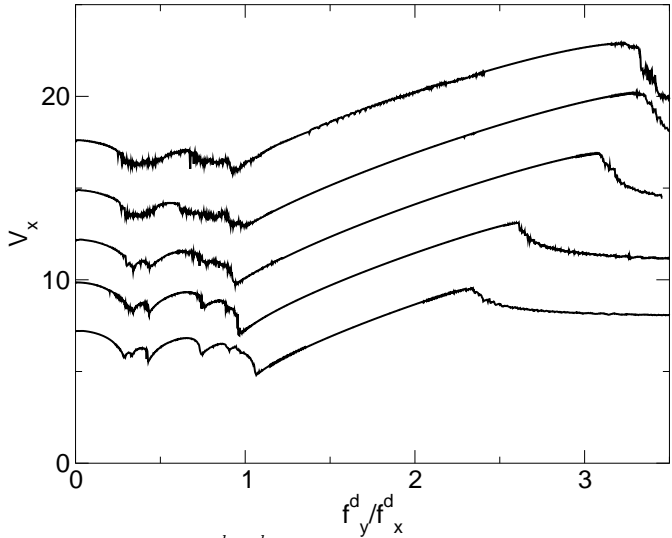}}
\caption{  
$V_{x}$ vs $f^{d}_{y}/f^{d}_{x}$ for the system in Fig.~9 with
a systematic shift added to the $y$ direction for clarity. From top
to bottom, $q_{d}/q =$ 2.0, 1.75, 1.5, 1.0, and 0.75. 
}
\end{figure}

\noindent
at $q_{d}/q = 1.75$. The
width of the locking regions for
$f_{y}^d/f_{x}^d < 1.1$ 
do not increase much with $q_d$; however, 
they became increasingly smeared,
which is particularly noticeable for $q_{d}/q \geq 1.75$.
The smearing occurs when
the lattice distortions become large enough 
to create dislocations in the surrounding lattice. These dislocations 
appear when the particle jumps from one symmetry angle
to another, and thus they smear out the transition.  The saturation of 
the width of the main $(1,0)$ step also coincides with the onset
of the creation of dislocations as the particle
passes into and out of the locked phase.
In Fig.~10 we plot the corresponding $V_{x}$ vs $f^{d}_{y}/f^{d}_{x}$ 
for the system in Fig.~9. Again we find a smearing of the
features with increasing $q_d$ for $f^{d}_{y}/f^{d}_{x} < 1.0$.
Additionally, we find an increase in the
magnitude of the fluctuations in $V_{x}$ just above
the end of the $(1,0)$ locking phase for $q_{d}/q = 2.0$,
due to the formation of dislocations in the 
surrounding media.  

In Fig.~11 we plot a
phase diagram showing the
evolution of the widths of the transverse pinned phase $T_p$ and
the $(1,0)$ locked phase $L$. The
phases are identified from the features in $V_{y}$ shown in Fig.~9. 
Here the transverse pinned phase $T_{p}$ grows with $q_d$
until its width saturates around
$q_{d}/q = 0.75$, while the locked phase L also grows with $q_d$ and
saturates at $q_d/q=1.75$. 
There is a small decrease in the width
of the locked phase for higher values of $q_{d}$. 
Above $q_{d}/q=2.5$, 
the phases become increasingly difficult to distinguish due to the 
large distortions in the lattice that cause
large velocity fluctuations near the
transitions into and out of the locked phases.  
  
The $(1,0)$ phase grows
with increasing particle interaction strength $q_d$
because the driven particle can more easily distort the surrounding lattice, 
allowing the lattice to accommodate the $(1,0)$ 
phase for larger angles between the driving force and $60^\circ$. 
Eventually, for 
large 

\begin{figure}
\center{
\epsfxsize=3.5in
\epsfbox{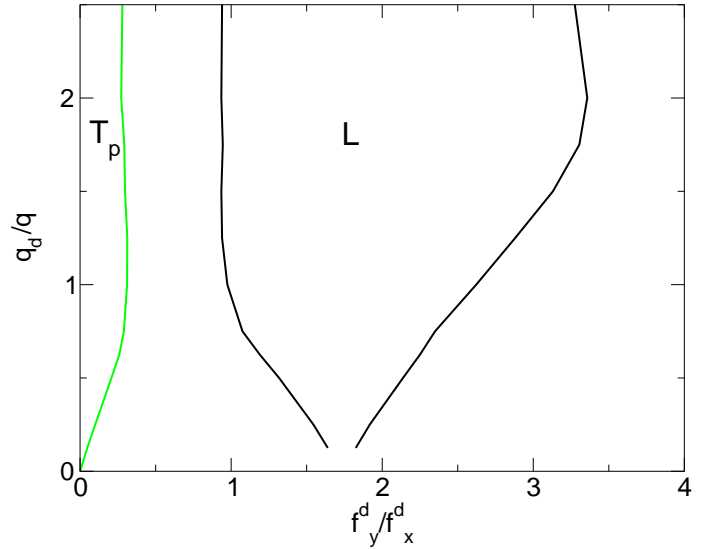}}
\caption{  
Phase diagram for $q_{d}/q$ vs $f^{d}_{y}/f^{d}_{x}$. 
The transverse pinned phase is marked $T_{p}$, and the
locked phase for $m = 1$, $n = 0$ is marked L. 
}
\end{figure}

\noindent
enough $q_{d}/q$, 
the distortion in the surrounding lattice become
strong enough to allow dislocations to be created,
enabling the particle to jump out of the locked channel,
and limiting the growth of the locked phase $L$ in Fig.~11. 
As the particle interaction increases
above $q_d/q=2$, this transition occurs at
lower values of $f^{d}_{y}/f^{d}_{x}$, and the 
width of the $(1,0)$ phase decreases 
as seen in Fig.~11.  

The behavior of the locked phase in Fig.~11 also indicates how this system
could be used for particle species separation. If 
two different species of particles are moving through the sample with
the same applied $f_{y}^d/f_{x}^d$, then it is possible for one of the
species to be in the locked L phase while the other species is not,
and thus the two species will move in different directions. For example, 
for a drive of $f_{y}^d/f_{x}^d = 1.25$, particles with $q_{d}/q > 0.75$
are in the L phase while particles with $q_{d}/q < 0.75$ are
not and will move at smaller angles. 
One advantage of this technique over the optical
traps is that particles with the same optical properties but different 
charges or sizes 

\begin{figure}
\center{
\epsfxsize=3.5in
\epsfbox{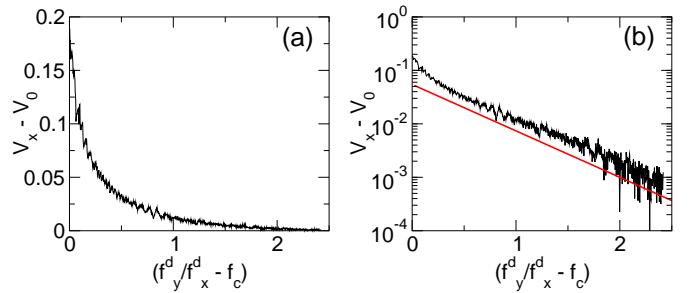}}
\caption{  
(a) $V_{x} - V_{0}$ vs $f^{d}_{y}/f^{d}_{x} - f_{c}$,
where $V_{0} = 1.0$ and 
$f_{c}=2.338$ is the value of the driving force when the $(1,0)$ locking 
phase ends for a system with $q_{d}/q = 0.75$. 
(b) Log-linear plot of (a). The solid straight line is a fit
with the form $A\exp(-bx)$ where $b = 2.0$ and $A = 0.1$. 
}
\end{figure}

\begin{figure}
\center{
\epsfxsize=3.5in
\epsfbox{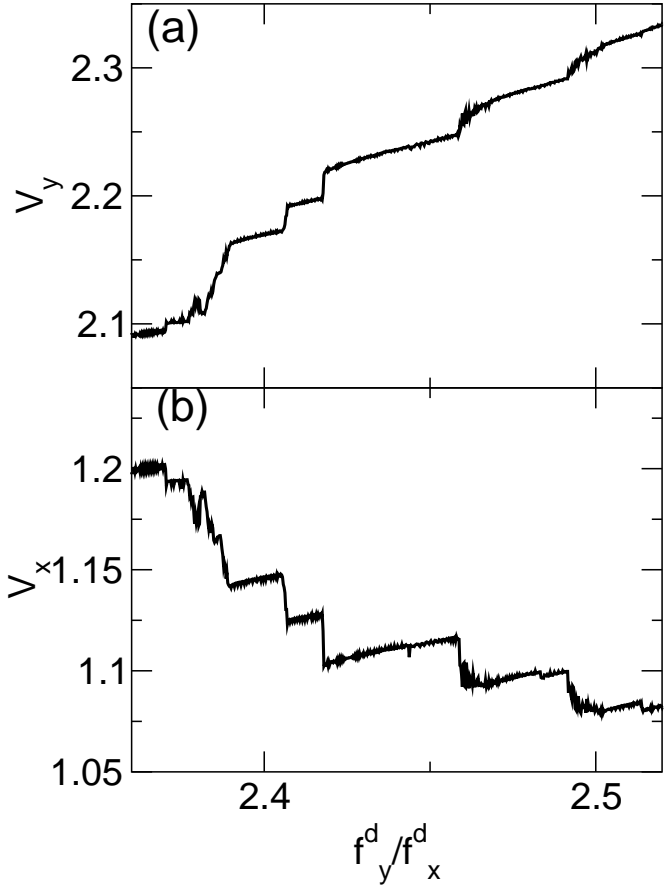}}
\caption{  
(a) $V_{y}$ vs $f^{d}_{y}/f^{d}_{x}$ for the system in Fig.~12,
just after the $(1,0)$ locking ends.  
(b) Corresponding $V_{x}$. 
}
\end{figure}

\noindent
could be separated.     

\subsection{Dynamics just outside of the $m= 1$, $n = 0$ phase} 

We now look more closely at some features in the $V_{x}$ and $V_{y}$ curves
as well as 
the particle dynamics just at the end of the $m = 1$, $n = 0$ phase.
In Fig.~10, after the end of the $(1,0)$ phase,
the $V_{x}$ curves decay back to $V_{x} = 1.0$.
All the values $q_{d}/q < 2.5$ show a similar decay in this regime. 
To examine this decay, in Fig.~12(a) we plot 
$V_{x} - V_{0}$ vs $f^{d}_{y}/f^{d}_{x} - f_{c}$, where 
$q_{d}/q = 0.75$, $V_{0} = 1.0$, and 
$f_{c}=2.338$ is the driving force at which the 
$(1,0)$  locked regime ends and
the $V_{x}$ decay begins. 
In Fig.~12(b) 
we plot the same curve in log-linear form and show a
good fit to an exponential
form $A\exp(-bx)$, with $b = 2.0$. For other
values of $q_{d}/q$ we observe the same behavior
with similar values for $b$. 
The fitting constants and curve shape do not change 
when the drive is swept more slowly
in the simulations, indicating that we are not observing a
transient behavior. 

In the decay region shown in Fig.~12, there appear to be some 
smaller oscillations in $V_{x}$.  Similar oscillations are
also seen in $V_{y}$. 
Focusing on this region, we find
that 

\begin{figure}
\center{
\epsfxsize=3.5in
\epsfbox{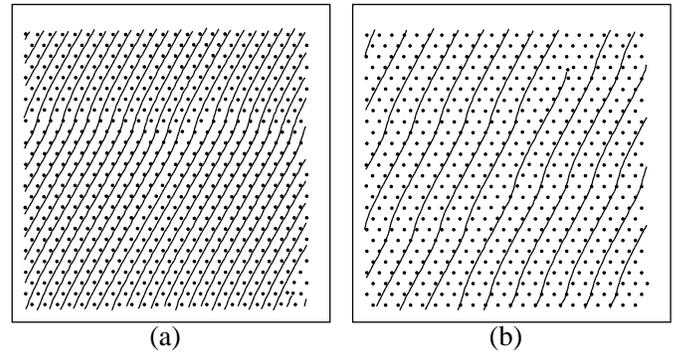}}
\caption{  
Particles (black dots) and trajectories (black lines) for
the system in Fig.~13 at:
(a) the step at $f^{d}_{y}/f^{d}_{x} = 2.412$, and 
(b) the step at $f^{d}_{y}/f^{d}_{x} = 2.51$. 
}
\end{figure}

\noindent
the oscillations are actually steps. In Fig.~13(a) we plot $V_{y}$
for the case of $q_{d}/q = 0.75$ right after the end of the 
$(1,0)$ locked phase.
Here $V_{y}$ does not increase smoothly but passes through a series of small
jumps. Between these jumps $V_{y}$ increases linearly. In Fig.~13(b)
we plot
the corresponding $V_{x}$ for the same range of driving forces. Here $V_{x}$
also  decreases in a
series of jumps which coincide with the jumps in $V_{y}$.  
Between the jumps, $V_{x}$ increases linearly.
For higher values of $f_{y}^d/f_{x}^d$ beyond what is shown 
in Fig.~13, the jumps continue 
and become more frequent until they overlap. 
In order to determine the cause of the steps, 
in Fig.~14 we plot 
the particle trajectories
on the small step centered at 
$f_{y}^d/f_{x}^d = 2.412$ from Fig.~13. 
Here the particle moves at $60^\circ$
for a distance of about $17a$, then 
jumps over to the next row, moves at $60^\circ$ again, and then repeats
the cycle of alternately jumping and moving.
In Fig.~14(b), on the step just after $f_{y}^d/f_{x}^d = 2.51$
a similar motion is seen as in Fig.~14(a); 
however, the particle moves a distance of about $9a$ before jumping 
to the adjacent row. This 
staircase type motion is similar to 
the dynamics just above the 
transverse depinning transition as shown in Fig.~6. In both these
cases the particle moves predominantly
along the symmetry angle which is $0^\circ$ for the transverse depinning
and $60^\circ$ here, with periodic jumps into the next row.  
For the higher order steps, the particle moves a shorter distance
in the $60^\circ$ direction before jumping to the next row. 

\section{Strongly Interacting Particles}
We next consider the case of driven particles
with $q_{d}/q > 2.5$. 
In this regime we find that a significant number of dislocations
are introduced into the system. In Fig.~15 we show $V_{y}$ and 
$V_{x}$ for $q_{d}/q = 5.0$. 
Here the curves are noisy with several
sharp steps. No clearly defined locking regime is present;
however, there are remnants of the $(1,0)$ locking regime.

The smaller phase locking regions for drive angles less than $60^\circ$
that were present at smaller $q_{d}$ are 
completely washed out in Fig.~15.
On these smallest steps, as 

\begin{figure}
\center{
\epsfxsize=3.5in
\epsfbox{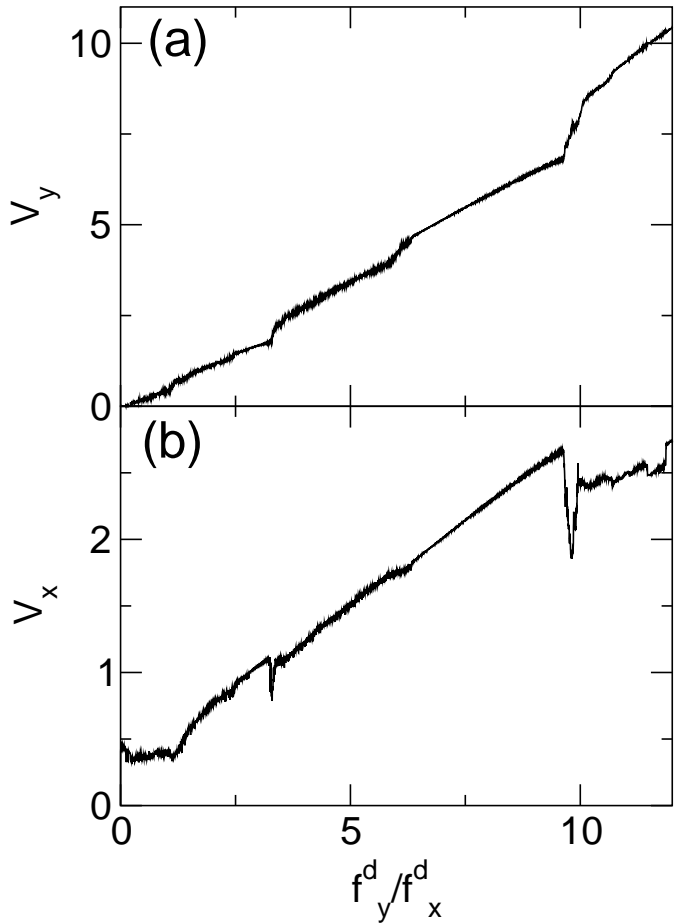}}
\caption{  
(a) $V_{y}$ vs $f^{d}_{y}/f^{d}_{x}$ for $q_d/q=5$.
(b) Corresponding $V_{x}$ vs $f^{d}_{y}/f^{d}_{x}$.
}
\end{figure}

\noindent
illustrated in Fig.~8,
the driven particle created the largest distortions in 
the surrounding lattice.  
As $q_{d}$ is increased, dislocations appear first 
for the steps which produce the
largest lattice distortions, and hence 
the smallest phase locking regions are smeared out first. 
This explains the disappearance of these phases 
at large $q_{d}$. Fig.~15 shows that there is still an appreciable 
locking effect for the $m = 1$, $n  = 0$ phase. The intermittent jumps
in the velocities 
are due to the formation of dislocations, a process which
abruptly changes the velocity of the driven particle.
Additionally, a small
number of dislocations remain in the colloid lattice for 
$f^{d}_{y}/f^{d}_{x}$ below the $m = 1$, $n = 0$ phase.      
In Fig.~16 we show the trajectories for the
system in Fig.~15   
for $f_{y}^d/f_{x}^d = 1.0$. 
Here the trajectories are strongly disordered and in general do
not overlap. 
For this value of drive a number of dislocations are created
in the lattice. Some of the
dislocations disappear when the lattice reorders once the 
driven particle moves past; 
however, some persist for long times. Since we are using periodic boundary
conditions, the particle passes through the same region repeatedly and
interacts with the persistent dislocations, 
which could make
the 

\begin{figure}
\center{
\epsfxsize=3.5in
\epsfbox{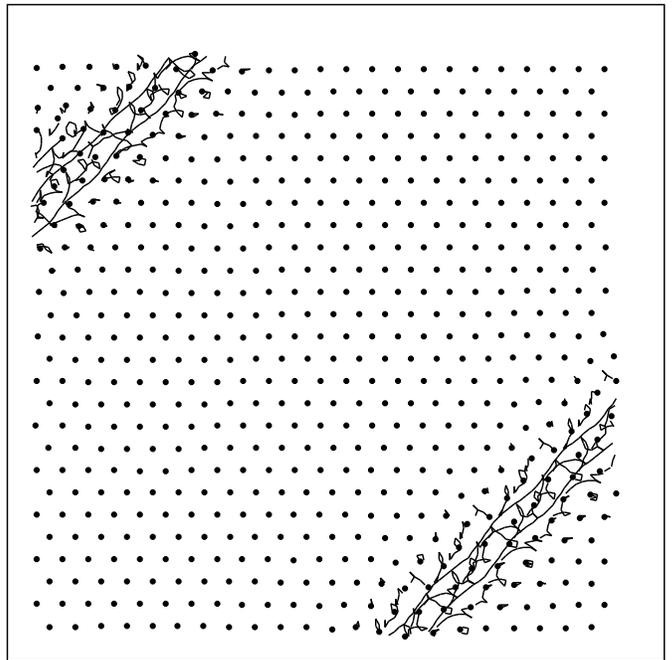}}
\caption{  
Particles (black dots) and trajectories (black lines) for the
system in Fig.~15 at $f^{d}_{y}/f^{d}_{x} = 1.0$.
}
\end{figure}

\noindent
flow more disordered. To address this we have 
performed the same simulations for
larger systems and examined the trajectories before the particle comes 
back on its previous path. We find the same disordered flow,
indicating that it is the process of dislocation creation which causes
the particle trajectories to become disordered, and not the
existence of dislocations from a previous passage.
We have also measured the power spectra in this case and find
white noise rather than a narrow band signal.

\section{Temperature Effects}
In a real colloidal system, temperature effects will be
relevant. For a triangular colloidal system at a finite
density, there is a well defined melting temperature $T_{m}$ above which
dislocations proliferate. We investigate the 
effects of the temperature on the locking up to $T_{m}$.
We concentrate on the system with $q_{d}/q = 0.5$ where
locking is still clearly visible but the 
driven particle does not generate dislocations in the
surrounding media at $T = 0$. In Fig.~17(a) we
show $V_{x}$ for $T/T_{m} = 0.0$, 0.44, 0.7, 0.9, and
$1.01$. For the lowest temperature, the width of the
$(1,0)$ step is the largest. It decreases for higher $T$
and disappears above $T_{m}$. Additionally,
the initial dip in $V_{x}$ for the transversely pinned phases is
also reduced as the temperature increases. The higher order locked regions 
for $f^{d}_{y}/f^{d}_{x} < 1.4$ become washed out 
for $T/T_{m} \ge 0.7$.

In Fig.~17(b) we plot the width $W$ of the      
$(1,0)$ locking step from Fig.~17(a). Here the width deceases
with increasing $T$ and drops suddenly at $T_{m}$. This
indicates that, even for relatively small particles, 
the locking effects 

\begin{figure}
\center{
\epsfxsize=3.5in
\epsfbox{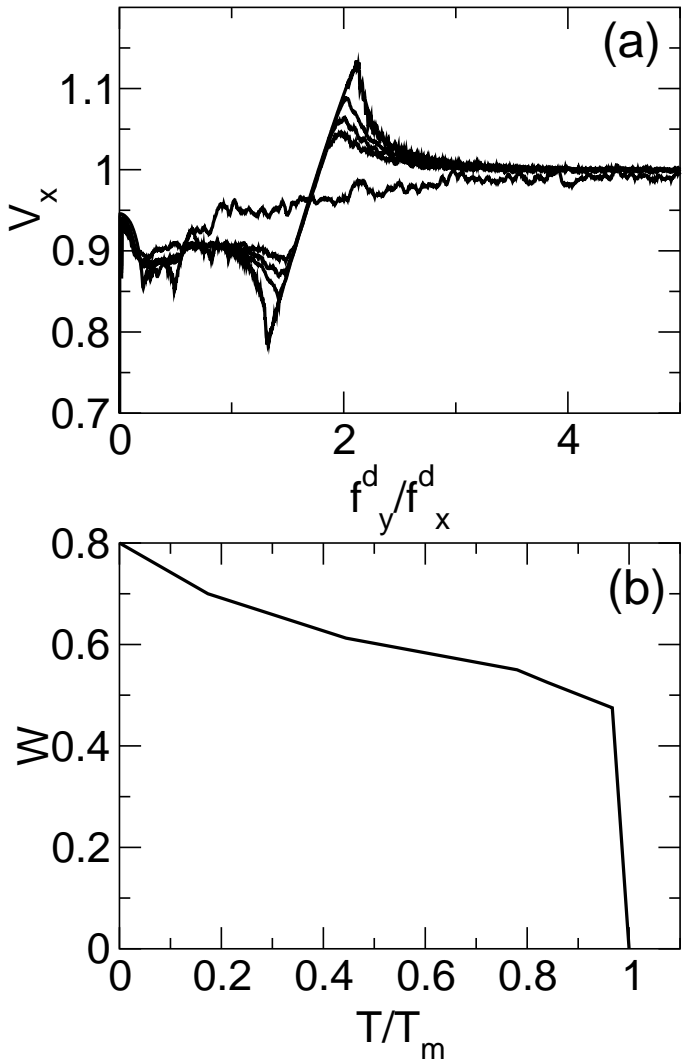}}
\caption{  
(a) $V_{x}$ vs $f^{d}_{y}/f^{d}_{x}$ for varied 
$T/T_{m} = 0.0$, 0.44, 0.7, 0.9, and 1.01. $T_{m}$ is the
temperature at which the non-driven lattice melts. 
The curve with the largest $(1,0)$ locking phase
(centered at $f^{d}_{y}/f^{d}_{x} = 1.7$) is $T = 0.0$. 
The width of the $(1,0)$ locking phase decreases for increasing $T$.
(b) Width $W$ of the $(1,0)$ locking vs $T/T_{m}$
for the system in Fig.~17(a). }
\end{figure}
\noindent
should be observable in an experimental system
all the way up to the melting transition of the colloidal
lattice itself.

\section{Summary}

In summary, we have investigated the dynamics of driven particles
moving through a triangular colloidal lattice for varied orientations
of the drive with respect to the symmetry of the colloidal lattice. 
This system, where
the driven particle causes distortions in the surrounding colloidal lattice, 
differs from previously studied systems in which
particles are driven through a lattice of {\it fixed} obstacles or traps. 
We find a series of locked phases as a function
of drive angle where the particles prefer to move along
certain symmetry directions. For driven particles that interact only 
weakly with the surrounding lattice, we find a Devil's staircase structure.
For larger driven particles, the
locking effects become much more pronounced and
the transitions into and out of the locked states
become much sharper. 
These transitions
can be accompanied by the creation of short-lived dislocations in the
surrounding lattice.
We also find new features such as negative differential resistance
in both the transverse and longitudinal velocities in the
intermediate particle regime. 
As the driven particles increase in size,
the smallest
locking regimes are lost when the main locking regimes grow and significant 
distortion of the surrounding lattice occurs. 
For larger particles,
the locking effects are washed out when significant numbers of dislocations
are created. 
Since the width of the locking steps depends strongly on the size of the
driven particle, our results may be useful as a method for
particle separation techniques or electrophoresis.   
We have also studied the dynamics 
at the transitions into and out of the main locking states
and find a staircase-like motion of the  driven particle.

We thank C. Bechinger, 
D.G.~Grier, M.B. Hastings, P.~Korda, X.S. Ling, A. Pertsinidis, and G. Spalding 
for useful discussions. 
This work was supported by the U.S. DoE under Contract No. W-7405-ENG-36.

\end{document}